\input epsf
\documentstyle[emulateapj]{article}
\begin{document}
\newcommand{\om}{\Omega_{\rm m}}
\newcommand{\ov}{\Omega_\Lambda}
\newcommand{\ob}{\Omega_{\rm b}}
\newcommand{\oq}{\Omega_{\rm Q}}
\newcommand{\Dnl}{\Delta_{\rm nl}}
\newcommand{\Dl}{\Delta_{\rm l}}
\newcommand{\LCDM}{\Lambda{\rm CDM}}
\def\go{\mathrel{\raise.3ex\hbox{$>$}\mkern-14mu
\lower0.6ex\hbox{$\sim$}}}
\def\lo{\mathrel{\raise.3ex\hbox{$<$}\mkern-14mu
\lower0.6ex\hbox{$\sim$}}}

%%%%%%%%%%%%%%%%%%%%%%%%%%%%%%%%%%%%%%%%%%%%%%%%%%%%%%%%%%%%%%%%%%%%%%%%%
\submitted{Submitted 1999 May 11; accepted 1999 June 9}
\title{The Mass Power Spectrum in Quintessence Cosmological Models }

\author{Chung-Pei Ma\footnote{Department of Physics and Astronomy,
                University of Pennsylvania, Philadelphia, PA 19104;
		cpma@strad.physics.upenn.edu},
        R.~R. Caldwell\footnote{Department of Physics,
                Princeton University, Princeton, NJ 08544},
	Paul Bode\footnote{Department of Astrophysical Sciences,
                Princeton University, Princeton, NJ 08544}, 
        and Limin Wang\footnote{Department of Physics,
                Columbia University, New York, NY 10027}}

%%%%%%%%%%%%%%%%%%%%%%%%%%%%%%%%%%%%%%%%%%%%%%%%%%%%%%%%%%%%%%%%%%%%%%%%%
\begin{abstract}
We present simple analytic approximations for the linear and fully
evolved nonlinear mass power spectrum for spatially flat cold dark
matter (CDM) cosmological models with quintessence (Q).  Quintessence
is a time evolving, spatially inhomogeneous energy component with
negative pressure and an equation of state $w_Q<0$.  It clusters
gravitationally on large length scales but remains smooth like the
cosmological constant on small length scales.  We show that the
clustering scale is determined by the Compton wavelength of the
Q-field and derive a shape parameter, $\Gamma_Q$, to characterize the
linear mass power spectrum.  The growth of linear perturbations as
functions of redshift, $w_Q$, and matter density $\Omega_m$ is also
quantified.  Calibrating to $N$-body simulations, we construct a
simple extension of the formula by Ma (1998) that closely approximates
the nonlinear power spectrum for a range of plausible QCDM models.
\end{abstract}


%%%%%%%%%%%%%%%%%%%%%%%%%%%%%%%%%%%%%%%%%%%%%%%%%%%%%%%%%%%%%%%%%%%%%%%%%
\section{Introduction}

Quintessence offers an alternative to the cosmological constant
($\Lambda$) as the missing energy in a spatially flat universe with a
sub-critical matter density $\om$ (Caldwell et al. 1998 and references
therein).  It is an energy component which, similar to $\Lambda$, has
negative pressure and therefore a negative $w_Q$ in the equation of
state $p_Q=w_Q\,\rho_Q\,$.  However, unlike $\Lambda$ for which $w=-1$,
quintessence is time evolving and spatially inhomogeneous, and $w_Q$
can have a range of values.  The observational imprints of the
quintessence therefore differ from those of the commonly studied
$\Lambda$CDM cosmology (e.g., Wang et al. 1999).

In this {\it Letter}, we study spatially flat QCDM models in which the
cold dark matter and Q-field together make up the critical density
(i.e. $\om+\Omega_Q=1$).  The quintessence is modelled as a scalar
field that evolves with a constant equation of state $w_Q$.  It drives
the cosmological expansion at late times, influencing the rate of
growth of structure.  Fluctuations in Q behave as an ultra-light mass
scalar field: on very large length scales the quintessence clusters
gravitationally, thereby modifying the level of cosmic microwave
background temperature anisotropy relative to the matter power
spectrum amplitude (in addition to a late-time integrated Sachs-Wolfe
effect); on small length scales, fluctuations in Q disperse
relativistically and the Q-field behaves as a smooth component.

We investigate the effects of the quintessence on the spectrum and
time evolution of gravitational clustering in both the linear and nonlinear
regimes.  We propose simple, analytical fitting formulas for both the
linear and fully evolved nonlinear power spectrum of matter density
fluctuations in plausible QCDM models.  For the linear power spectrum
(\S~2), we introduce a simple parameter $\Gamma_Q$ derived from the
Compton wavelength of the Q-field to characterize its shape.  This
parameter determines the length scale above which the Q-field can
cluster gravitationally, and is reminiscent of $\Gamma_\nu$ derived
from the free streaming distance of hot neutrinos in cold+hot dark
matter (C+HDM) models by Ma (1996).  For the nonlinear power spectrum
(\S~3), we examine the validity of the simple linear to nonlinear
mapping technique that has been successfully developed for scale-free,
CDM, C+HDM, and $\LCDM$ models (Hamilton et al. 1991; Jain, Mo, \&
White 1995; Peacock \& Dodds 1996; Ma 1998).  We present a simple
extension of the analytical formula of Ma (1998) that closely
approximates the QCDM nonlinear power spectrum computed from a set of
$N$-body simulations.

The formulas presented in this {\it Letter} are essential for gaining
physical insight into the effects of the quintessence on gravitational
collapse and for performing rapid predictions of observable quantities
in the linear as well as nonlinear regimes in plausible QCDM models.

%%%%%%%%%%%%%%%%%%%%%%%%%%%%%%%%%%%%%%%%%%%%%%%%%%%%%%%%%%%%%%%%%%%%%%%%%
\section{Linear Power Spectrum}

We use the conventional form to express the linear power
spectrum\footnote{Defined so that the two-point correlation function
is $\xi(r)\equiv 4\pi \int k^2 dk \,P(k) \sin(k r)/(k r)$.}  for the
matter density perturbation $\delta_m$ in QCDM models:
\begin{equation}
	P(k,a) = A_Q\, k^n\, T^2_Q(k) \,\left( 
	{a\,g_Q\over g_{Q,0}} \right)^2 \,,
\label{pl}
\end{equation}
where $A_Q$ is a normalization, $k$ is the wavenumber, $n$ is the
spectral index of the primordial adiabatic density perturbations, and
$T_Q$ is the transfer function which encapsulates modifications to the
primordial power-law spectrum.  The function $g_Q$ is the linear
growth suppression factor, $g_Q=D/a$, where $D$ is the standard linear
growth factor for the matter density field in QCDM models, and
$g_{Q,0}=g_Q(a=1)$ denotes its value at the present day with scale
factor $a=1$.  We discuss each piece of equation (\ref{pl}) in turn.

\vspace{-.2in}
\centerline{\vbox{\epsfxsize=11.5cm\epsfbox{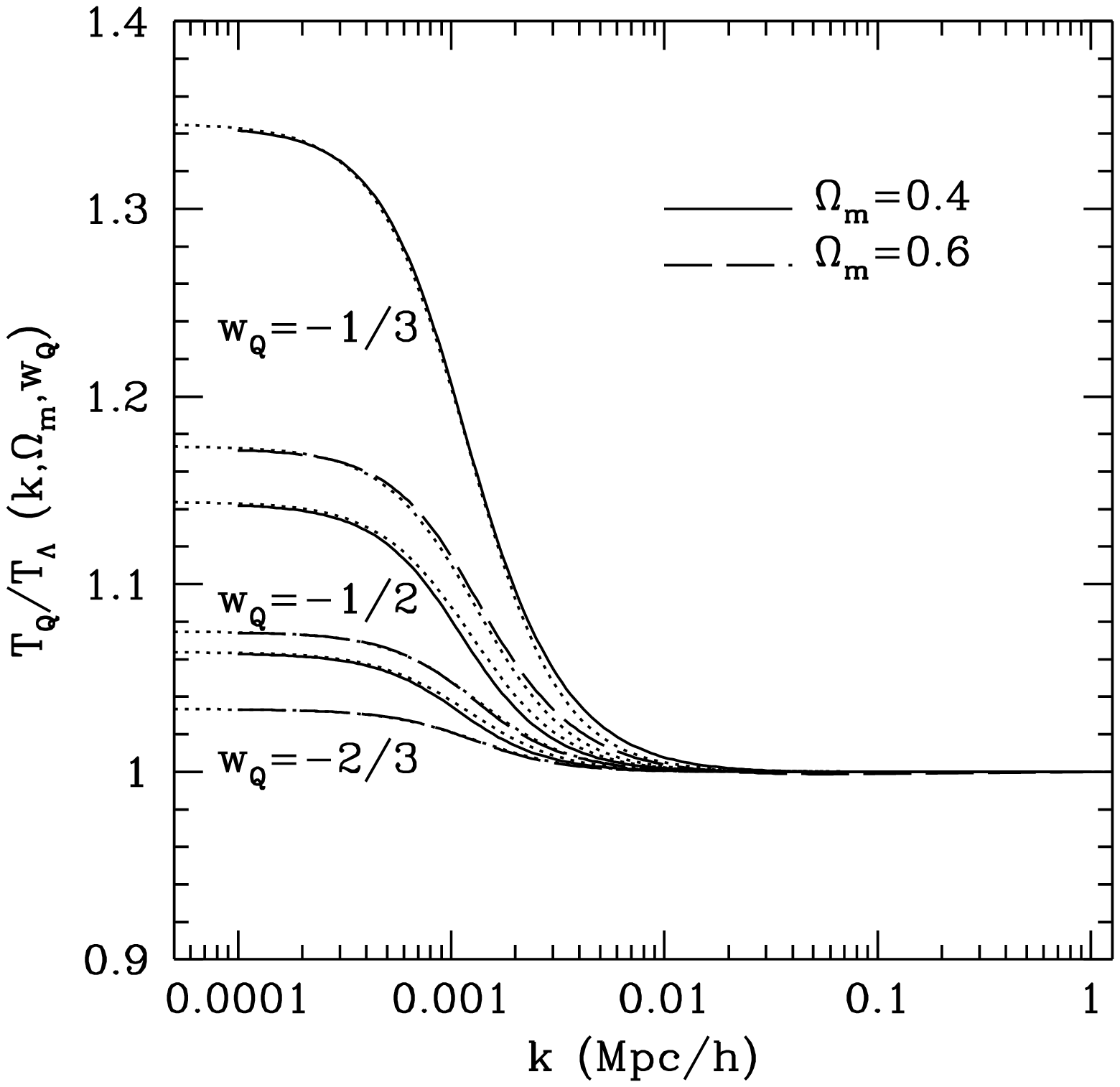}}}
\vspace{-.3in}
\noindent{\small
FIG.~1 $-$ Ratio of the transfer functions, $T_{Q\Lambda}\equiv
T_Q/T_\Lambda$, at the present day for six pairs of flat QCDM and
$\LCDM$ models.  The solid (for $\om=0.4$) and dashed ($\om=0.6$)
curves are computed from the Boltzmann integrations and illustrate the
dependence on the matter density parameter $\om$.  For a given $\om$,
the three curves illustrate the dependence on the equation of state:
$w_Q=-1/3, -1/2$, and $-2/3$ from top down.  The dotted curves show
the analytic approximation given by equation~(\ref{kq})-(\ref{alpha}).
Note that $T_{Q\Lambda}$ deviates from unity only on very large length
scale ($k\lesssim 0.01\,h$ Mpc$^{-1}$) above which the Q-field can
cluster spatially.}

\vspace{.3in}
First we examine the transfer function for the matter density field.
To isolate the effects of quintessence, we find it convenient and
illuminating to compare a pair of QCDM and $\LCDM$ models that have
the same set of cosmological parameters and differ only in $w_Q$
(recall $w=-1$ for $\LCDM$).  We define the relative transfer function
for a pair of such models to be $T_{Q\Lambda} = T_Q/T_\Lambda$.  For
$T_\Lambda$, we follow the convention and set the arbitrary amplitude
of $T_\Lambda$ to unity as $k \rightarrow 0$. The form of $T_\Lambda$
is well known and various fitting formulas have been published (e.g.,
Bardeen et al. 1986; Efstathiou, Bond, \& White 1992; Sugiyama 1995).
More complicated fits with higher accuracy have also been developed
for higher baryon ratios ($\ob/\om\go 20\%$) and for the features due
to baryonic oscillations and damping (e.g., Bunn \& White 1997;
Eisenstein \& Hu 1998).

The transfer function $T_Q$ for QCDM models resembles $T_\Lambda$ for
$\LCDM$ but with one key difference.  The linear matter density field,
$\delta_m$, evolves according to the equation $\ddot\delta_m + 2
H\dot\delta_m = 4\pi G (\rho_m\delta_m +\delta\rho_Q + 3\,\delta
p_Q)$, where $H=\dot a/a$ and the dots denote differentiation with
respect to proper time.  On small length scales, the Q-field is smooth
($\delta\rho_Q,\,\delta p_Q \ll \delta\rho_m$) and we recover the
familiar equation for the evolution of $\delta_m$ (Caldwell et
al. 1998).  On very large length scales, however, the Q-field clusters
and contributes to the energy density and pressure perturbations.  The
result is a different growth rate for $\delta_m$ on large and small
scales once the quintessence starts to dominate the cosmological
energy density.  We can determine the characteristic scale separating
these two regimes by examining the linear equation for the $Q$-field:
$\ddot{\delta Q} + 3 H\dot{\delta Q} + (k^2 + V_{,QQ})\delta Q
=\dot\delta_m [(1 + w_Q)\rho_Q]^{1/2}$ (where $V$ is the Q-field
potential, $V_{,QQ}\equiv d^2V/dQ^2$, and $\rho_Q=\dot{Q}^2/2+V$).  We
see that $\delta Q$ itself behaves as a scalar field with an effective
mass $(V_{,QQ})^{1/2}$ and a Compton wavenumber of $k_Q\sim
(V_{,QQ})^{1/2}$.  On small length scales (i.e. $k\gg k_Q$), the
amplitude of $\delta Q$ and hence $\delta\rho_Q$ is damped and does
not enter the evolution equation for $\delta_m$.  On large scales
($k\ll k_Q$) $\delta Q$ grows, so the Q-field clusters and in turn
affects the evolution of $\delta_m$.

The change in the behavior of $\delta_m$ near $k \sim (V_{,QQ})^{1/2}$
as a result of differing $Q$-clustering properties is illustrated in
Figure~1 for $T_{Q\Lambda}$ vs. $k$ for a range of $w_Q$ and
$\om$.  We have chosen to normalize $T_{Q\Lambda}$ to unity at the
high-$k$ end because both the Q-field and the cosmological constant
are spatially smooth on these scales.  The clustering property of Q is
reminiscent of the case of massive neutrinos in C+HDM models, which
cannot cluster appreciably below the neutrino free-streaming scale but
can cluster with the same amplitude as the cold dark matter on large
scales.  Analogous to the shape parameter $\Gamma_\nu$ that was
introduced to model the neutrino streaming distances in C+HDM models
(Ma 1996), we introduce a new shape parameter $\Gamma_Q$ here to
characterize the feature in Figure~1 in QCDM models.  For a
constant equation of state, $w_Q$, one can show that $V_{,QQ} = 6\pi G
(1 - w_Q) (2\rho + p + w_{Q}\rho)\,$, where $\rho$ and $p$ are the
total energy density and pressure.  We approximate
\begin{eqnarray}
	k_Q \! &=& \! \Gamma_Q\,h = 2\sqrt{V_{,QQ}} \nonumber\\
	\! &=& \! {3 H\over c}\sqrt{(1-w_Q)[2 + 2 w_Q - w_Q \om(a)]}
\label{kq}
\end{eqnarray}
and we use a simple ratio of polynomials to express the relative
transfer function:
\begin{equation}
	T_{Q\Lambda}(k,a) \equiv {T_Q\over T_{\Lambda}} =
	{\alpha + \alpha\,q^2 \over 1 + \alpha\, q^2}\,, \qquad 
	q={k\over \Gamma_Q\,h}\,,
\label{tql}
\end{equation}
where $k$ is in Mpc$^{-1}$, and $\alpha$ is a scale-independent but
time-dependent coefficient that quantifies the relative amplitude of
the matter density field $\delta_m$ on large and small length scales.
We find $\alpha$ well approximated by
\begin{eqnarray}
       \alpha \!  &=& \! (-w_Q)^s \,, \label{alpha} \\ 
  	 s \! &=& \! (0.012 - 0.036\,w_Q - 0.017/w_Q) [1-\om(a)]  \nonumber\\
	\! &+& \! (0.098 + 0.029\,w_Q - 0.085/w_Q) \ln\om(a) \,, \nonumber
\end{eqnarray}
where the matter density parameter is $\om(a) =
\om/[\om + (1 - \om)\, a^{-3 w_Q}]\,$, which reaches the value $\om$ at
the present day $a=1$.  Figure~1 illustrates the close
agreement (with errors $\lesssim 10\%$) between the approximations
given by equations~(\ref{kq}) - (\ref{alpha}) and the exact results
from numerical integrations of the Boltzmann equations.
 
\centerline{\vbox{\epsfxsize=11.5cm\epsfbox{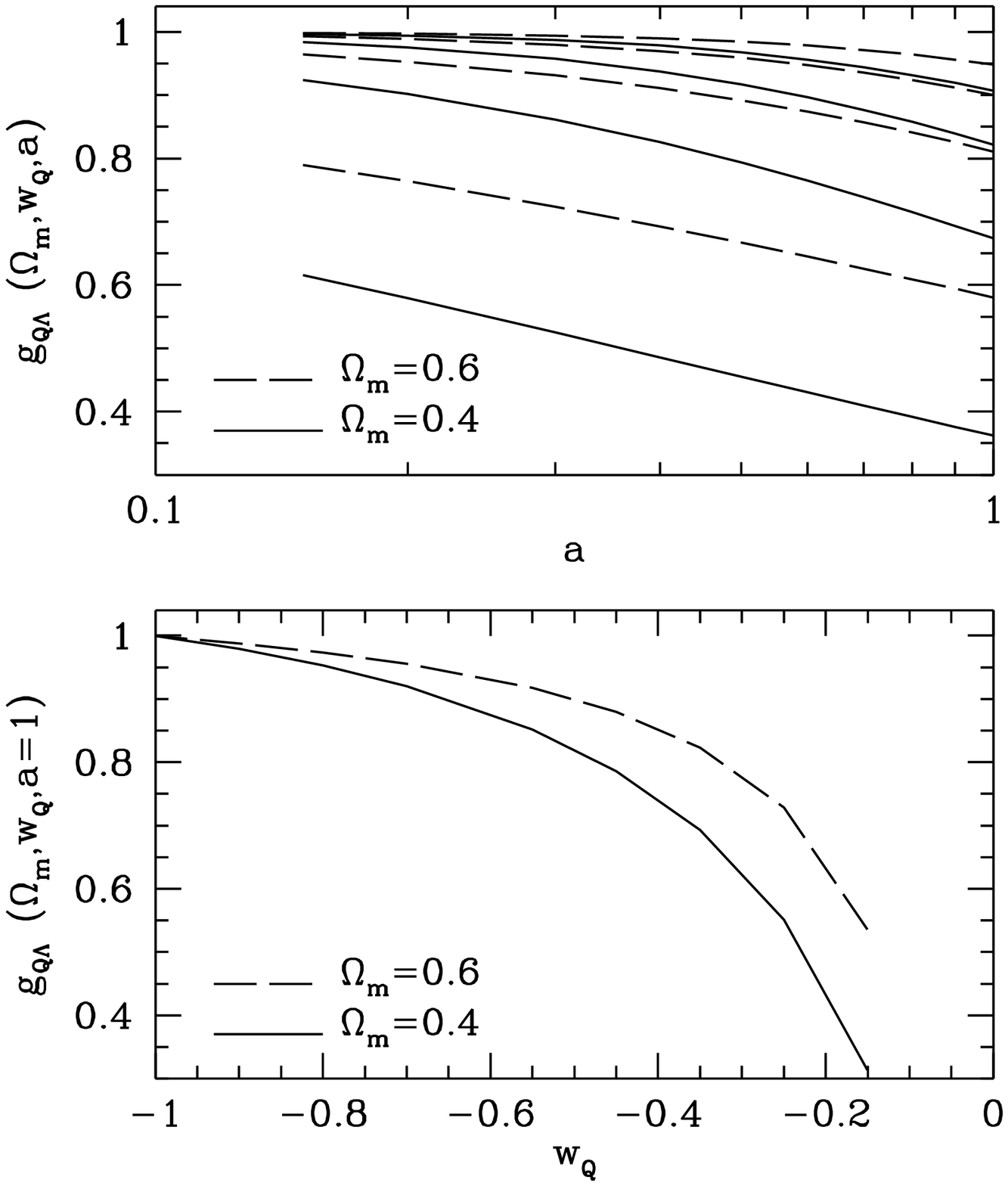}}}
\noindent{\small
FIG.~2 $-$ Ratio of the growth suppression factors, $g_{Q\Lambda} \equiv
g_Q/g_\Lambda$, as a function of the scale factor $a$ (top) and the
equation of state $w_Q$ (bottom; at $a=1$) for various pairs of flat
QCDM and $\LCDM$ models.  The solid and dashed curves in both panels
are for $\om=0.4$ and 0.6, respectively.  In the top panel, each set
of curves corresponds to $w_Q=-2/3, -1/2, -1/3$, and $-1/6$ from top
down.}

\vspace{.3in}
Next we examine the linear growth suppression factor of the density
field in equation~(\ref{pl}).  This function is well studied for
$\LCDM$ models (Heath 1977; Lahav et al. 1991).  An empirical fit is
given by $g_\Lambda = 2.5 \om(a) \{\om(a)^{4/7}- 1 + \om(a) +
[1+\om(a)/2] [1 + (1-\om(a))/ 70] \}^{-1}$ and is accurate to $\sim
2\%$ for $0.1 \le \om \le 1$ (Carroll, Press, \& Turner 1992).  This
formula unfortunately cannot be generalized to QCDM models by simply
replacing $(1-\om)\rightarrow (1-\om) a^{-3(1+w_Q)}$.  Instead, we
propose the following formula to approximate the ratio of the QCDM and
$\LCDM$ growth factors:
\begin{eqnarray}
  g_{Q\Lambda} \! &\equiv & \! {g_Q \over g_\Lambda} = (-w_Q)^t\,,
	\label{gql}\\
    t \! & = & \! -(0.255 + 0.305\,w_Q + 0.0027/w_Q) [1-\om(a)]\nonumber\\
      \! & - & \! (0.366 + 0.266\,w_Q - 0.07/w_Q) \ln\om(a) \, \nonumber
\end{eqnarray}
accurate to $2\%$ for $0.2 \lo \om \le 1$ and $-1 \le w_Q \lo -0.2$.
Figure~2 illustrates the dependence of $g_{Q\Lambda}$ on
time, $w_Q$, and $\om$.  The growth is evidently slower in models with
less negative $w_Q$ for fixed $\om$.  This is because the energy
density in the Q-field dominates over that in matter at an
increasingly earlier time as $w_Q$ is varied from $-1$ to 0; the
growth of gravitational collapse therefore ceases earlier and results
in a smaller value for $g_{Q\Lambda}$.  (It is sometimes useful to
study the instantaneous growth rate of $\delta_m$, $f\equiv
d\log\delta_m /d\log\,a$.  See Wang \& Steinhardt (1998) for a fitting
formula for $f$.)

The remaining component in equation~(\ref{pl}) to be specified is the
normalization $A_Q$.  It can be chosen by fixing the value of
$\sigma_8$, the rms linear mass fluctuation within a top hat of radius
$8\,h^{-1}$ Mpc, or by fixing to COBE results.  For the latter we
follow Bunn \& White (1997).  In the case that the temperature
anisotropy is due to primordial adiabatic density perturbations with
spectral index $n$, we write $A_Q=\delta_H^2 (c/H_0)^{n+3}/(4\pi)\,$,
where $\delta_H =2 \times 10^{-5}\alpha_0^{-1}\,(\om)^{c_1 + c_2
\ln\om}\exp{[c_3(n-1) + c_4(n-1)^2]}\,$, $\alpha_0=\alpha(a=1)$ of
equation~(\ref{alpha}), $c_1= -0.789 |w_Q|^{0.0754 - 0.211 \ln
|w_Q|}\,$, $c_2= -0.118 - 0.0727 w_Q\,$, $c_3=-1.037$, and
$c_4=-0.138$, for $-1\lo w_Q \lo -0.2$.  In the case of tensor
perturbations, the primordial amplitudes follow the inflationary
relation $A_T = 8(1 - n) A_S\,$.  The ratio of $\ell=10$ multipole
moments, $r_{10} = C_{10}^T/C_{10}^S$, is given by $r_{10}\approx
0.48(1-n)[1 + 0.1 (1-n) {(8 + 7 w_{Q}){\oq}^2 + 3]} (1 -
\oq/x)^{g_{10}(\oq/x)}\,$, where $g_{10}(y) = 0.18 + 0.84 y^2\,$, and
$x = 0.75[1-0.66 w_Q + 1.66 w_Q^2 - 0.5 (1+w_Q)^5]\,$.  One can
rescale $A_Q$ by $A_Q \to A_Q/(1+r_{10})$ to accommodate the effect of
tensors on the normalization.

%%%%%%%%%%%%%%%%%%%%%%%%%%%%%%%%%%%%%%%%%%%%%%%%%%%%%%%%%%%%%%%%%%%%%%%%%
\section{Non-Linear Mass Power Spectrum}

In this section we examine if the simple linear to nonlinear mapping
technique initiated by Hamilton et al. (1991) can be extended to QCDM
models.  The basic approach is to search for a simple expression for
the function $\Dnl(k) = f[\Dl(k_0)]$ that relates the linear and
nonlinear density variance $\Delta(k) \equiv 4\pi k^3 P(k)$.  Note
that $\Dnl$ and $\Dl$ are evaluated at different wavenumbers, where
$k_0=k(1+\Dnl)^{-1/3}$ corresponds to the precollapsed scale of $k$.
The strategy is to combine analytical clustering properties in
asymptotic regimes with fits to numerical simulation results.  This
recipe has been successfully developed for scale-free models with a
power-law $P(k)$ (Hamilton et al. 1991; Jain et al. 1995), flat CDM
and $\LCDM$ models (Jain et al. 1995; Peacock \& Dodds 1996, PD96
hereafter; Ma 1998), and flat C+HDM models with massive neutrinos (Ma
1998, Ma98 hereafter).

We investigate if the PD96 and Ma98 formulas proposed for $\LCDM$
models can be easily extended to QCDM models.  These two formulas
incorporate the time dependence of the mapping in different ways, but
they share the feature that the dependence on parameters $\om$ and
$\ov$ enters only through the linear growth factor $g$.  In order to
test the application of this method to QCDM models, we have performed
$N$-body simulations for three values of $w_Q$: $-2/3, -1/2$, and
$-1/3$, each with several different realizations.  These three values
should be sufficient since extensive tests of $w_Q=-1$ (i.e.  $\LCDM$
models) have already been carried out in PD96 and Ma98.  We restrict
our attention to $w_Q < -1/3$ and cosmological parameter ranges
that are in concordance with observations (Wang \& Steinhardt 1998;
Wang et al. 1999).  Specifically, $(\om,\oq,\ob,h)= (0.4, 0.6, 0.047,
0.65)$ for the $w_Q=-2/3$ and $-1/2$ models, and
$(\om,\oq,\ob,h)=(0.45, 0.55, 0.047, 0.65)$ for the $w_Q=-1/3$ model.
The $N$-body code used is a parallel version of the particle-particle
particle-mesh algorithm (Bertschinger \& Gelb 1991; Ferrell \&
Bertschinger 1994).  Each simulation uses $128^3$ particles in a box
of comoving volume $100^3$~Mpc$^3$.  The Plummer softening length is
50~kpc comoving, which allows us to compute the nonlinear power
spectrum in highly clustered regions with $k\lo 10\,h$~Mpc$^{-1}$ and
$\Dnl \lo 1000$.  Since the Q-field clusters only on scales much above
the box size, the presence of the quintessence only affects the
initial conditions and the evolution of the scale factor $a$.

\vspace{.2in}
\centerline{\vbox{\epsfxsize=11.cm\epsfbox{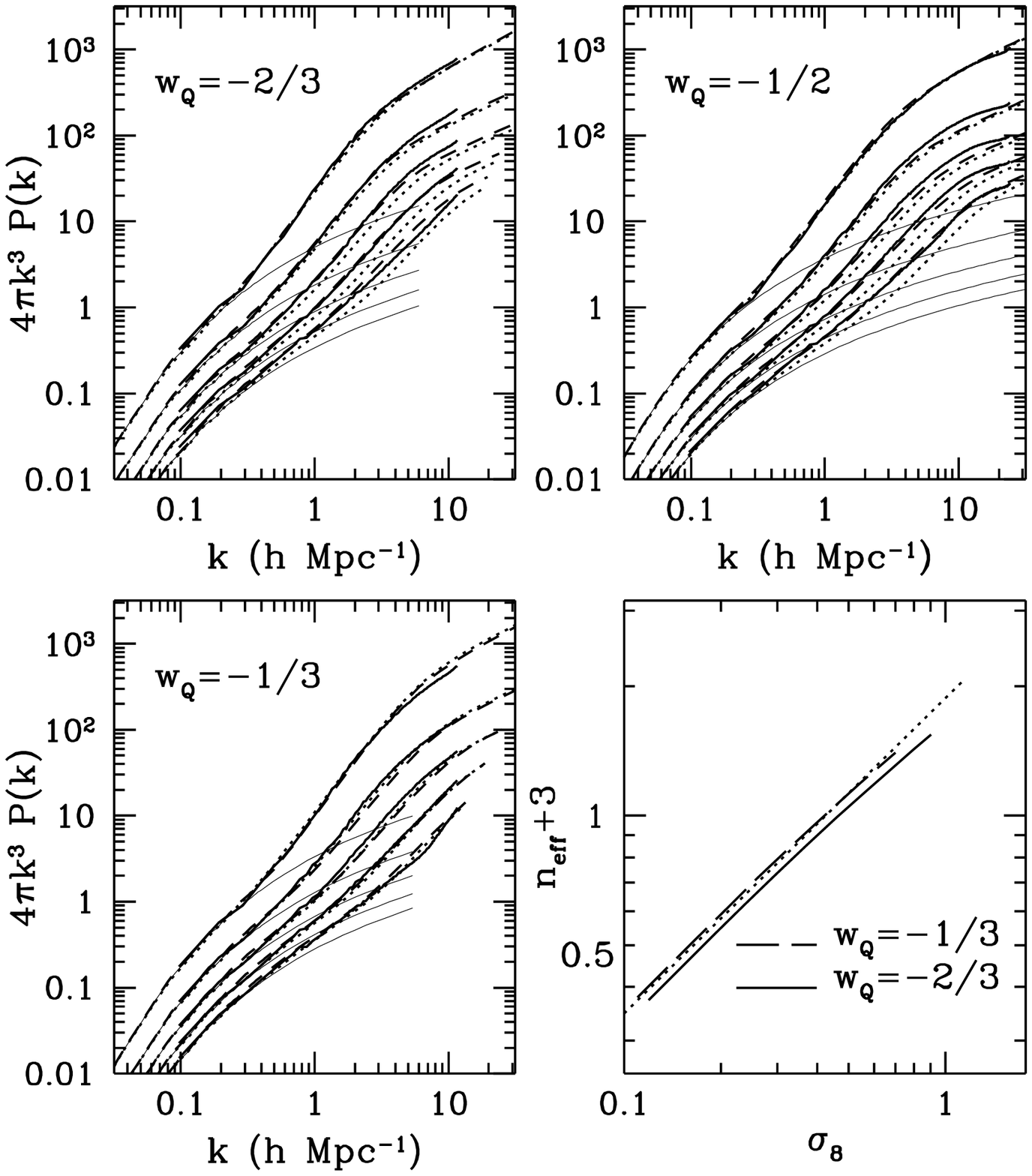}}}
\noindent{\small
FIG.~3 $-$ {\it Lower left and upper}: The linear and fully evolved 
nonlinear power spectra for the matter density field in QCDM models
with different equations of state $w_Q$ (see text for other model
parameters).  In each panel, five redshifts, $z=0$, 1, 2, 3, and 4,
are shown (top down).  The curves are computed from: $N$-body
simulations directly (thick solid), nonlinear approximation by Ma98
(dashed; eq.~(6)) and PD96 (dotted), and linear theory (thin solid).
{\it Lower right}: The effective spectral index, $n_{\rm eff}$, as a
function of $\sigma_8$ for two QCDM models.  The dotted line
represents a power law and demonstrates that $d\ln(n_{\rm
eff}+3)/d\ln\sigma_8 \propto \beta$ is an excellent approximation.}

\vspace{.3in}
Figure~3 compares the linear power spectrum and the fully
evolved spectrum from both the $N$-body runs and the approximations of
PD96 and Ma98.  Five redshifts are shown for each of three QCDM
models.  Overall, we find that the PD96 formula works well at $z=0$
when the factor $g$ in their formula is set to $g=g_Q$, which is the
appropriate growth factor for the density field for QCDM models
[eq.~(\ref{gql})].  At earlier times, however, the PD96 formula {\it
underestimates} $\Dnl$ at $k\go 1\,h$ Mpc$^{-1}$ in the $w_Q=-2/3$ and
$-1/2$ models by up to 30\%.  We have attempted less
physically-motived combinations of growth factors (e.g.,
$g=\alpha\,g_Q$ and $g_\Lambda$) but did not find a way to make PD96
fit.  We find that the Ma98 formula,
\[
	\frac{\Dnl(k)}{\Dl(k_0)} = 
	G \left( \frac{\Dl}{g_0^{3/2} \sigma_8^\beta} \right)\,,
\]
\begin{equation}
	G(x) = [1 + \ln(1+ 0.5\,x )]\,{1 + 0.02\,x^4 + c_1\,x^8 /g^3 \over
	1 + c_2\, x^{7.5}}\,, 
\label{dnl}
\end{equation}
%\end{eqnarray}
can be easily extended to QCDM models.  Specifically, we propose to
keep $c_1= 1.08\times 10^{-4}$ and $c_2 = 2.10 \times 10^{-5}$ used
for $\LCDM$ in Ma98, but adopt $g=g_Q\,$, which is the appropriate QCDM
growth factor [eq.~(\ref{gql})], and
$g_0=|w_Q|^{1.3\,|w_Q|-0.76}\,g_{Q,0}\,$, where $g_{Q,0}\equiv
g_Q(a=1)$.  As described in Ma98, the parameter $\beta$ in
equation~(\ref{dnl}) is introduced to approximate the power-law dependence
of the effective spectral index $n_{\rm eff}+3$ in previous work on
$\sigma_8$: $d\ln(n_{\rm eff}+3)/d\ln\sigma_8\propto \beta\,$.  We find
$\beta=0.83$ an excellent approximation for all three QCDM models that
we tested (see the lower-right panel of Figure~3). Other panels of
Figure~3 illustrate the close agreement (rms errors $\sim 10$\%)
between $N$-body results and equation~(\ref{dnl}).

%%%%%%%%%%%%%%%%%%%%%%%%%%%%%%%%%%%%%%%%%%%%%%%%%%%%%%%%%%%%%%%%%%%%%%%%%
\vspace{.2in}

\section{Summary}

We have presented simple formulas to approximate both the linear and
nonlinear power spectra for matter density perturbations in viable
quintessence cosmological models with an equation of state $-1\le w_Q
\lo -1/3$.  Equations~(\ref{kq}),(\ref{tql}), and (\ref{alpha}) together
specify the ratio of the linear transfer functions $T_Q$ and
$T_\Lambda$ for the matter density field for a given pair of QCDM and
$\LCDM$ models with the same cosmological parameters.
Equation~(\ref{gql}) specifies the ratio of the linear growth
suppression factors $g_Q$ and $g_{\Lambda}$ in QCDM and $\LCDM$ models.
Equation~(\ref{dnl}) approximates the nonlinear mass power spectrum.

A key difference between gravitational clustering in QCDM and $\LCDM$
models is that $\Lambda$ is spatially smooth on all length scales,
whereas the Q-field can cluster above a certain length scale.  We
characterize this length scale by the shape parameter $\Gamma_Q$ of
equation~(\ref{kq}), which is derived from the Compton wavelength of
the Q-field.  The QCDM matter power spectrum therefore changes shape
at two characteristic scales: $\Gamma_Q$, and the familiar $\Gamma
\propto \om\,h$ that corresponds to the cross-over from radiation- to
matter-dominated era.  For the QCDM models studied in this
{\it Letter} (i.e. constant $w_Q$), the Compton wavelength of the
Q-field is very large: $k_Q \sim 0.001$ to $0.01\,h$ Mpc$^{-1}$.  On
scales of galaxy clusters and below, therefore, the linear QCDM power
spectrum has {\it identical} shape as in the corresponding $\LCDM$
model and differs only in the overall amplitude by a factor of
$(A_Q/A_\Lambda) (g_{Q\Lambda}/g_{Q\Lambda,0})^2$.  This realization
should simplify comparisons between QCDM and $\LCDM$ models.

For the fully evolved nonlinear power spectrum, we find that PD96
works well at $z=0$ but underestimates its amplitude by
up to $\sim 30$\% at earlier times.  The formula of Ma98, on the other
hand, can be easily extended to approximate the QCDM nonlinear $P(k)$
(with errors $\lesssim 10$\%) for $w_Q\lo -1/3$ and redshift up to
$z\approx 4$.  Equation~(\ref{dnl}) summarizes this result.

\vspace{.5in}
%%%%%%%%%%%%%%%%%%%%%%%%%%%%%%%%%%%%%%%%%%%%%%%%%%%%%%%%%%%%%%%%%%%%%%%%%
Supercomputing time was provided by the National Scalable Cluster
Project at UPenn.  C.-P. M. acknowledges support from an Alfred P.
Sloan Foundation Fellowship, a Cottrell Scholars Award from the
Research Corporation, and a Penn Research Foundation Award.  Support
is also provided by DOE grants DE-FG02-91ER40671 (R.~C.),
DE-FG02-92ER40699 (L.~W.), and NSF grants AST-9318185/9803137,
ACI-9619019 (P.~B.).
%%%%%%%%%%%%%%%%%%%%%%%%%%%%%%%%%%%%%%%%%%%%%%%%%%%%%%%%%%%%%%%%%%%%%%%%%

\end{document}